# Pressure-induced phase transitions of multiferroic BiFeO$_3$[*]

ZHANG Xiao-Li(张晓丽) [1], WU Ye(吴也)[2], ZHANG Qian(张倩)[2], DONG Jun-Cai(董俊才)[1], WU Xiang(巫翔)[2,2)], LIU Jing(刘景)[1], WU Zi-Yu(吴自玉)[1,3,3)], and CHEN Dong-Liang(陈栋梁)[1,1)]

[1] Beijing Synchrotron Radiation Facility, Institute of High Energy Physics, Chinese Academy of Sciences, Beijing 100049, People's Republic of China

[2] Key Laboratory of Orogenic Belts and Crustal Evolution, MOE, School of Earth and Space Sciences, Peking University, Beijing 100871, People's Republic of China

[3] National Synchrotron Radiation Laboratory, University of Science and Technology of China, Hefei, Anhui 230029, People's Republic of China

**Abstract**: Pressure-induced phase transitions of multiferroic BiFeO$_3$ have been investigated using synchrotron radiation X-ray diffraction with diamond anvil cell technique at room temperature. Present experimental data clearly show that rhombohedral (*R3c*) phase of BiFeO$_3$ first transforms to monoclinic (*C2/m*) phase at 7 GPa, then to orthorhombic (*Pnma*) phase at 11 GPa, which is consistent with recent theoretical *ab initio* calculation. However, we observe another peak at $2\theta=7°$ in the pressure range of 5-7 GPa that has not been reported previously. Further analysis reveals that this reflection peak is attributed to the orthorhombic (*Pbam*) phase, indicating the coexisting of monoclinic phase with orthorhombic phase in low pressure range.

**PACS**: 61.50.Ks，64.70.K-,

[*]Supported the National Natural Science Foundation of China under Grant No. 10979060 and 11275220.
1) Email: chendl@ihep.ac.cn
2) Email: xiang.wu@pku.edu.cn
3) Email: wuzy@ihep.ac.cn

# 1.INTRODUCTION

Multiferroics, which exhibit both ferroelectric and magnetic order in the same phase, have attracted great interests because of these unique properties[1, 2]. Especially, the magnetic properties of multiferroics can be modulated by external electric fields and vice versa. Bismuth ferrite, $BiFeO_3$ (BFO) as a prototype multiferroic material, is perhaps the only compound with both magnetic and ferroelectric at ambient conditions[3, 4]. In BFO, the $6s^2$ lone pair electrons of $Bi^{3+}$ arouse the ferroelectricity until the temperature reaches $T_C$=1100 K[5]. Spiral spin cycloid modulated G-type antiferromagnetic order with a periodicity of 62 nm occurs below the Neel temperature ($T_N$=640K)[5, 6]. Due to these promising physical properties, many experimental and theoretical efforts have been devoted to its structural characteristics[7] and the nature of magnetoelectric coupling[8, 9].

At ambient conditions, BFO crystallizes in the rhombohedral structure with *R3c* (Z=161) space group, with *c*-axis parallel to the diagonals of the perovskite cube[10]. The $Fe^{3+}$ ion is coordinated by six oxygen atoms, forming an octahedral $FeO_6$. Research of phase transitions under high pressures of BFO has been of long-standing interest[3, 7, 11, 12]. Early literature has reported that no phase transition occurs in BFO until the high pressure reaches up to 70 GPa[13]. In contrast, some studies have demonstrated that BFO has undergone several structural phase transitions during the upstroke process. At high pressure above 10 GPa, Haumont *et al*.[12] and Guennou *et al*.[14] found that BFO transforms to the orthorhombic *Pnma* structure instead of others. While at low pressure, BFO exhibits a variety of phase transitions. For instance, by using x-ray diffraction (XRD) and far-infrared spectroscopy, Haumont *et al*.[12] found that BFO remains the rhombohedral structure up to 6.2 GPa, then changes to a monoclinic structure with space group *C2/m*. Nevertheless, in a more recent study, a phase transition from the polar rhombohedral *R3c* phase to the antipolar orthorhombic *Pbam* phase with antiferroelectric character of atomic displacements has been revealed at 3 GPa by using neutron powder diffraction[15]. Furthermore, Guennou *et al*. [14] have reported that three different phase transitions of orthorhombic symmetry occur in BFO below 10 GPa, through Raman spectroscopy and XRD measurements. Besides, these three intermediate phases may be unstable, competing and coupling with each other. In a word, the crystal structural transitions of BFO at high pressure above 10 GPa are very clear, while at low pressure range below 10 GPa they are still remaining controversial.

In order to address the phase transitions of BFO under pressure, especially the structure behavior below 10 GPa, we have investigated its phase transition sequence at high pressure up to 53 GPa using *in-situ* synchrotron radiation x-ray diffraction. We find that at low pressure of 5-7 GPa, the monoclinic and orthorhombic phases coexist in BFO, competing with each other. Afterwards, the monoclinic phase becomes predominant and then transforms to the nonpolar orthorhombic phase.

# 2.EXPERIMENTAL DETAILS

BFO polycrystalline powder was prepared by sol-gel method[16] using Bismuth nitrate pentahydrate, and iron nitrate nonahydrate as starting materials in stoichiometric proportion. In present high-pressure experiments, asymmetric cell equipped with a pair of diamond anvil with the culet diameter of 300 μm was applied to compress the sample. BiFeO$_3$ powder firstly was compressed to a slice of ~20-μm thickness by tungsten carbide anvils, and then loaded into the chamber (110μm in diameter and 35 μm thickness) in a rhenium gasket. A ruby sphere was also loaded for pressure calibration [17]. Silicone oil was used as pressure-transmitting medium (PTM) to attain nearly quasi-hydrostatic compression of the sample for XRD.

*In situ* high-pressure XRD experiments were conducted at 4W2 beamline of Beijing Synchrotron Radiation (BSRF) by angle-dispersive measurements with a wavelength of 0.6199Å. The x-ray beam was focused in the horizontal and vertical direction to 30×7μm$^2$ (FWHM) spot. Powder diffraction patterns were collected with a Mar345 image plate and integrated with the ESRF Fit2D software package[18], followed by the GSAS[19] and FULLPROF[20] analyzing.

## 3. RESULTS AND DISCUSSION

### 3.1 Low-pressure region: 0-11GPa

The *in situ* high-pressure XRD data of BFO were collected at room temperature up to 53 GPa. Some selected XRD patterns at pressures below 11 GPa are plotted in Fig. 1. Observed reflections below 5 GPa are in good agreement with those of the rhombohedral *R3c* phase of BFO[10]. Upon compression, diffraction peaks of the rhombohedral BFO shift to higher angles and become broadening, leading to some peaks overlapping. All lattice constants of BFO at various pressures are refined by a full-profile model refinement (Le Bail method) and are displayed in Fig.2. At ambient pressure, the lattice parameters of the rhombohedral structure obtained from refinement are close to previously reported results[21], $a = b = 5.5756$ Å, $c = 13.8611$ Å. A fit of the pseudocubic volume in the stability region of the *R3c* phase with a third-order Birch-Murnaghan equation[22] leads to $V_0 = 62.20$ Å$^3$ and $K_T = 100(6)$ GPa for $K'$ being fixed to 4. This value compares reasonably with previously published values[15]. A monotonic decrease of the lattice parameters with the pressure increasing are clearly showed in Fig.2, which is attributed to the change in oxygen octahedral tilting[23].

Significant changes in XRD patterns appear above 5 GPa (arrows in Fig. 1), indicating the onset of the phase transitions. The first transition from *R3c* to phase II begins at 5 GPa evidenced by the emergence of new superstructure reflections. As mentioned previously [12], most of the diffraction peaks can be indexed with the monoclinic superstructure of *C2/m*. Above 7 GPa, this monoclinic *C2/m* structure becomes the dominant role in the pattern (Fig. 3), indicated by the splitting of (012) diffraction peak. The parameters at 7.1 GPa obtained from refinement are shown in Fig 3. Further increasing the pressure, volume of this phase tends to decrease linear resulted by pressure-induced shortening of the interatomic distance (Fig.2). These results are in good agreement with those reported by Haumont *et al*.[12]. The parameters obtained by fitting the *P-V* up to 11 Gpa with a third order Birch-Murnaghan equation of state using a pseudocubic cell are $V_0 = 61.20$ Å$^3$ and $K_T = 160(23)$ GPa for $K'$ being fixed to 4.

However, it is worth noting that at 5 GPa a weak but inneglible peak emergences at the low angle (2$\theta$=7°), accompanied by the occurrence of the first phase transition. To the best of our knowledge, this has not been reported up to now. We first suppose that it belongs to the orthorhombic (*Pbam*) structure, which has been reported at 3 GPa in BFO[15]. Refinement based on two models of monoclinic $2\sqrt{5}a_p \times 2a_p \times \sqrt{2}a_p$ spuerstructure of *C2/m* symmetry and orthorhombic $\sqrt{2}a_p \times 2\sqrt{2}a_p \times 2a_p$ superstructure of *Pbam* symmetry are employed to confirm the phase at 5.4 GPa. The best quality of fit shows that the observed reflections can be alternatively indexed using these two models (Fig. 4), i.e., the weak peak actually comes from the *Pbam* structure, revealing definitly that two phases co-exist in a certain narrow pressure range of 5-7 GPa. Since the crystals display large unit cells and complex domain structure, this maybe attributed to the competition between complex octahedral tilts and possible (anti)ferroelectric off-center cation displacement[12] in BFO, leading to the richness of phase transitions. Furthermore, two-phase coexistence also confirms the complex phases of perovskite structures under pressure or doping, just like the result that $PbZr_{1-x}Ti_xO_3$ (PZT) has confronted with a possible phase coexistence and/or micro-twinning[24].

**3.2 High-pressure region: 11-53 GPa**

In the high-pressure region above 11 GPa, we observe two apparent phase transitions as shown in Fig.5. The phase transition from *C2/m* to *Pnma* (phase III) occurs at 11 GPa, indicated by the appearance of reflection peak of $2\theta=13°$ (insets of Fig. 5). The nonpolar *Pnma* phase becomes predominant in the XRD pattern at 15 GPa, evidenced by the emergence of (200) and (210) diffraction peaks (Fig. 6), implying a distortion resulted by octahedra tilts. The parameters obtained from refinement are shown in Fig. 6. In this phase, the structure exhibits peak splitting due to a four-variant domain structure (insets of Fig. 5), which is commonly observed in *Pnma* crystals[14]. The values of the bulk modulus $K_T = 222(9)$ GPa for $K'$ being fixed to 4, while the unit cell volume per formula unit extrapolated to zero pressure leading to $V_0 = 60.01$ Å$^3$.

At high pressure of 33.5 GPa, the (111) peak of *Pnma* structure disappears while a new peak occurs at 15°. From refinement we find that the phase has already transformed to *Pnmm* structure[14] in which the new peak can be indexed to (401) (Fig. 7). Insets are the parameters of 42 GPa obtained from refinement. The parameters obtained by fitting the *P-V* up to 11 GPa with a third order Birch-Murnaghan equation of state using a pseudocubic cell are $V_0 = 55.01$ Å$^3$ and $K_T = 270(25)$ GPa for $K'$ being fixed to 4. Nevertheless, $a_R$ and $b_R$ show no appreciable change during this transition, confirming the group-subgroup relationships between *Pnma* and *Pnmm*. Afterwards, the cell volume remains decreasing, coincident with the decrease in the radius of ions in the transition from the high-spin state to the low-spin state[25], which may be associated with the transition from the insulating to a metallic state[26].

## 4. CONCLUSIONS

In summary, our *in-situ* high pressure XRD characterization of polycrystalline BFO reveals that it undergoes three phase transitions up to 53 GPa at room temperature. In contrast with previous XRD and Raman experimental results, we find that there exist unprecedented phase transitions of rhombohedral to monoclinic/orthorhombic structures at 5-7 GPa, much more complex than theoretical prediction[27]. Actually, two phases of monoclinic and orthorhombic coexist and compete with each other at this pressure range, due to the coupling between octahedral tilts and (anti)ferroelectric cation displacement. Our results reveal experimentally the structural richness of BFO, which has been proposed by Dieguez *et al.*[28] using first-principle calculations. With further increasing pressure the octahedral tilt is reduced while cation displacements are suppressed, leading to the nonpolar *Pnma* structure, which is in agreement with recent report.

**Acknowledgements**

The authors sincerely thank Lin Chuanlong for useful discussion.

Figure and tables caption

Fig. 1. Integrated diffraction patterns of the powder BFO for low-pressure below 11.3 GPa.

Fig. 2. Evolution of the lattice parameters (top) and volume (bottom) data of BFO. All parameters and volume data are normalized to the pseudo-cubic unit cell. The full pressure range is divided into five different regions.

Fig. 3. Powder x-ray diffraction profiles of 7.1 GPa of monoclinic *C2/m* structure. The observed profiles (crosses) and the difference (blue line) between the observed and fitted pattern (red line) are shown on the same scale. Insets are the parameters of the corresponding pressure deduced form the refinement. Vertical bars represent the calculated positions of diffraction peaks of *C2/m* structure.

Fig. 4. Powder x-ray diffraction profiles of 5.4 GPa for multiple phases of monoclinic and orthorhombic: Vertical bars represent the calculated positions of diffraction peaks of *Pbam* (olive) and *C2/m* (magenta) structures.

Fig. 5: Integrated diffraction patterns of the powder BFO for high-pressure (11-52 GPa) experiment.

Fig. 6. Powder x-ray diffraction profiles of 15 GPa of orthorhombic *Pnma* structure.

Fig. 7. Powder x-ray diffraction profiles of 42 GPa for *Pnmm* structure in BFO.

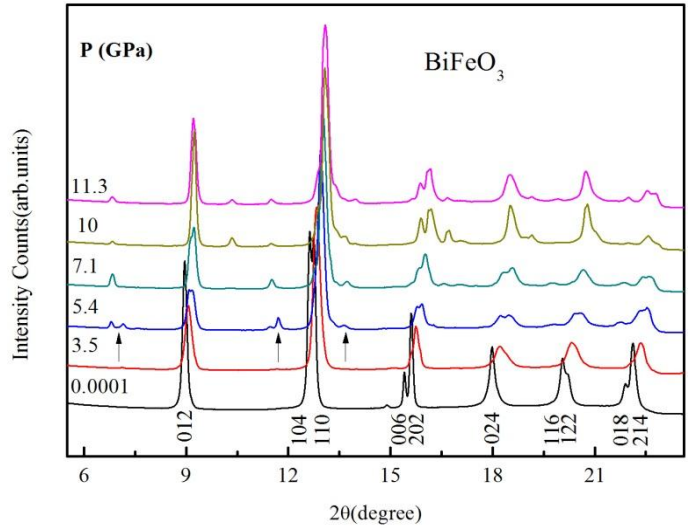

Fig. 1

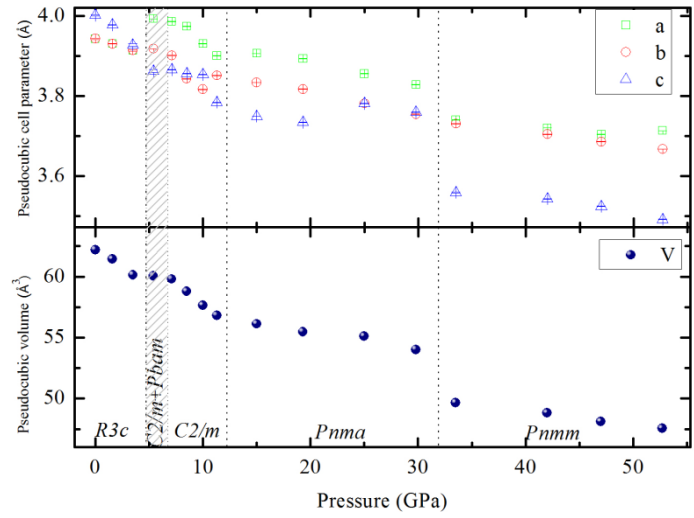

Fig. 2

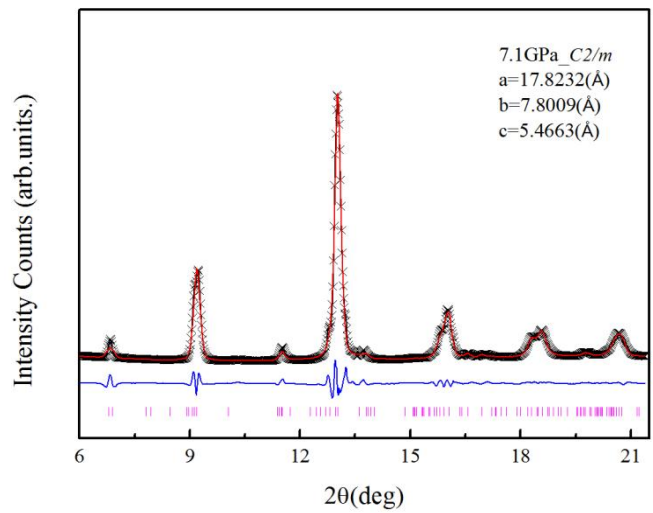

Fig. 3

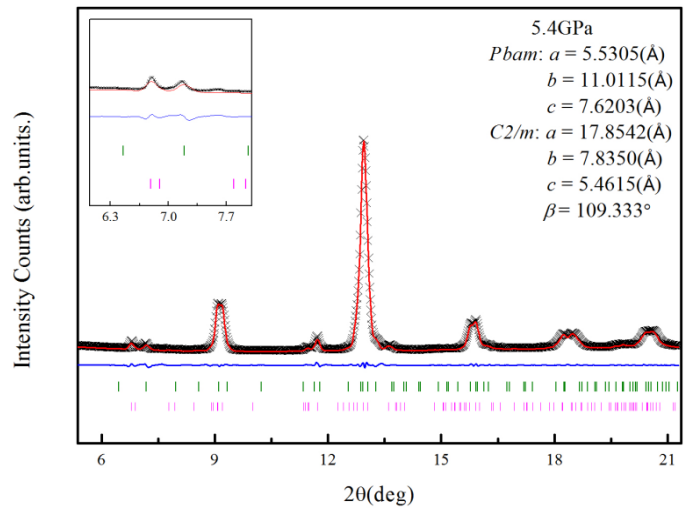

Fig. 4

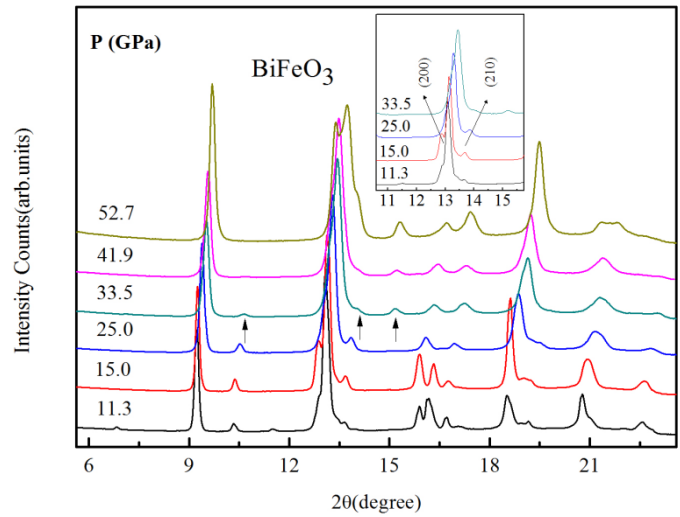

Fig. 5

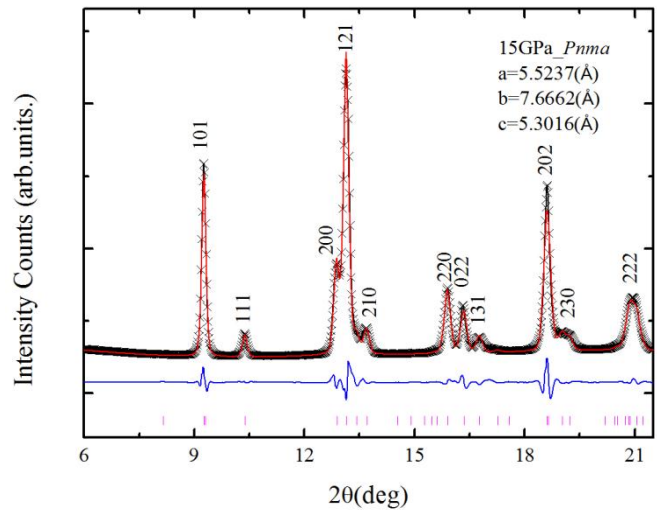

Fig. 6

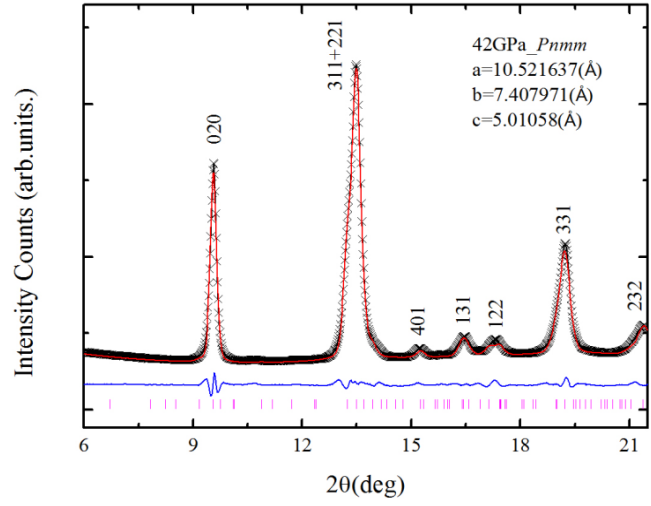

Fig. 7